\documentclass[runningheads]{llncs}
\usepackage[T1]{fontenc}
\usepackage{graphicx}
\usepackage[hidelinks]{hyperref}
\usepackage{dblfloatfix}
\usepackage{graphicx}
\usepackage{booktabs}
\usepackage{lipsum}
\usepackage{xcolor}
\usepackage{tabularx}
\usepackage[capitalise]{cleveref}
\usepackage{array}
\usepackage{cite}

\usepackage{amssymb}

\usepackage{multirow}
\usepackage{colortbl}
\usepackage{makecell}

\begin{document}
%

% \title{The Interest is Mutual: Enabling Multi-Stakeholder Job Recommendation through Heterogeneous Graph Neural Networks}

\title{OKRA: an Explainable, Heterogeneous, Multi-Stakeholder Job Recommender System}

\titlerunning{Multi-Stakeholder Explainable Job Recommendation}
% \titlerunning{Urban Bias in Job Recommendation}


% \author{Anonymous Author(s)}

\author{Roan Schellingerhout\orcidID{0000-0002-7388-309X} \and
Francesco Barile\orcidID{0000-0003-4083-8222} \and
Nava Tintarev\orcidID{0000-0002-5007-5161}}

\authorrunning{R. Schellingerhout et al.}

% \authorrunning{Anonymous Author(s)}


\institute{Maastricht University, Maastricht, Limburg, 6229 EN, The Netherlands
\email{\{roan.schellingerhout,f.barile,n.tintarev\}@maastrichtuniversity.nl}}
% \email{f.barile@maastrichtuniversity.nl}
% \email{n.tintarev@maastrichtuniversity.nl}

\maketitle              % typeset the header of the contribution
\begin{abstract}
  The use of recommender systems in the recruitment domain has been labeled as `high-risk' in recent legislation. As a result, strict requirements regarding explainability and fairness have been put in place to ensure proper treatment of all involved stakeholders. To allow for stakeholder-specific explainability, while also handling highly heterogeneous recruitment data, we propose a novel explainable multi-stakeholder job recommender system using graph neural networks: the \textit{Occupational Knowledge-based Recommender using Attention} (OKRA). The proposed method is capable of providing \textit{both} candidate- and company-side recommendations and explanations. We find that OKRA performs substantially better than six baselines in terms of nDCG for two datasets. Furthermore, we find that the tested models show a bias toward candidates and vacancies located in urban areas. Overall, our findings suggest that OKRA provides a balance between accuracy, explainability, and fairness. 

\end{abstract}

\section{Introduction}
%A common way for job seekers to find employment is through their social network \cite{wanberg2020job}. However, finding a job can become an insurmountable task for those with a limited network. Due to the enormous amount of possibilities %, all in different fields and requiring different skills, 
Finding a job can be compared to finding a needle in a haystack. As a result, the staffing industry has seen continuous growth over the past decades \cite{statista_2022}. Yet, even for experienced recruiters, successfully matching thousands of candidates to thousands of job vacancies can be prohibitively challenging. To assist in this task, recruitment agencies have increasingly been making use of machine learning to match candidates and job vacancies \cite{lavi2021consult,le2019towards,qin2018enhancing,qin2020enhanced,yildirim2021bideepfm,zhu2018person}. Within the field of recruitment, there exist several unique challenges and restrictions related to using machine learning. For one, the use of AI in recruitment has been labeled as `high-risk' in recent legislation \cite{euaiact2024}, due to the sensitive nature of the data being used and the large impact decisions can have on job seekers (candidates) and companies. This also highlights the requirement to consider the multiple stakeholders of recruitment, primarily \textit{candidates} - individuals looking for a job; \textit{recruiters} - those who try to match candidates to vacancies; and \textit{company representatives} - those who are responsible for hiring new employees in companies. 

Since different stakeholders within a multi-stakeholder environment often have conflicting goals, it is crucial that they can be presented with some justification of the provided recommendation \cite{abdollahpouri2020multistakeholder}. Without a properly tailored explanation, accepting recommendations at face value can be difficult. Despite explainability being a legal requirement for high-risk domains like recruitment \cite{euaiact2024}, the majority of state-of-the-art (SOTA) job recommender systems (JRSs) function as black boxes, meaning that their decisions are not explainable \cite{guo2020DeText,he2021self,kumari2023siamese,lavi2021consult,zhu2018person}; those that do offer some level of explainability tend to do so in a limited manner, without accounting for all stakeholders \cite{le2019towards,Upadhyay2021,yildirim2021bideepfm}. 

Furthermore, those same SOTA models opt to either make use of structured data (tabular data including skills, previous jobs, etc.) \cite{leksin2016job,yildirim2021bideepfm}, or unstructured data (CV and vacancy texts) \cite{guo2020DeText,he2021self,kumari2023siamese,lavi2021consult,le2019towards,zhu2018person}, failing to capitalize on both. Regardless of which data type is chosen, a portion of the data is left unused, potentially leaving a significant amount of predictive power underutilized. 

To address these shortcomings, this paper aims to answer the following research question: \textit{How can we design an explainable multi-stakeholder recommender system that outperforms state-of-the-art systems in user- and provider-side metrics?} To assist in doing so, we formulate the following three sub-questions:

\begin{description}
    \item[SQ1:] How well do state-of-the-art baseline models perform ranking on our datasets of structured and unstructured data?
    \item[SQ2:] How can we develop an explainable graph neural network that is able to fully utilize heterogeneous data to outperform state-of-the-art baselines?
    \item[SQ3:] How does such a model perform regarding user- and provider-side fairness metrics compared to state-of-the-art baselines?
\end{description}

%Our results show that both simple and state-of-the-art baselines struggle to perform well on our dataset, likely due to the task's difficulty. Considering the high level of detail required to distinguish texts within recruitment, current SOTA models often fall short. 
We find that combining structured and unstructured data leads to a considerable improvement in ranking accuracy (nDCG). Additionally, taking a stakeholder-specific approach can lead to further improvement, while also providing explainability. Furthermore, we find that all models, both text- and graph-based, perform somewhat poorly regarding user- and provider-side fairness metrics, meaning that they unfairly discriminate against candidates and companies from rural areas, demonstrating an urban bias.
\section{Related Work and Hypotheses}
The European Union has labeled the use of AI in recruitment as a high-risk scenario, emphasizing the need for careful consideration of how job recommender systems are developed and used \cite{euaiact2024}. This concern arises from the significant impact employment decisions have on individuals' lives and the sensitive nature of the data these systems handle. 
The field of job recommendation has garnered a significant amount of interest in recent years, with advances in natural language processing (NLP) playing a pivotal role in enhancing the performance of JRSs. %However, the field has yet to converge on a single best approach due to the diverse nature of job datasets, which can vary greatly in terms of features, language, and data availability. 
As a result, previous work largely focuses on creating embeddings from CVs and vacancies and attempting to match those two \cite{guo2020DeText,lavi2021consult}. However, when only focusing on one type of data, whether structured or unstructured, a significant amount of data is left unused. % CVs and vacancies are often not exhaustive, leaving details out for the sake of brevity and succinctness; on the other hand, structured data, like skill sets and previous job titles, are often less expressive, including a wider range of data, but with fewer details on each item. 
Furthermore, while some previous research has included aspects like fairness and explainability in their approach \cite{le2019towards,Upadhyay2021,yildirim2021bideepfm}, these are often not the main focus of the models. %While incorporating explainability or fairness, even in a limited amount, is always beneficial, in a highly sensitive domain like recruitment, such concepts should arguably be at the forefront of the model design. 

\subsection{Job Recommender Systems}
Most previous work on job recommendations makes use of NLP-related techniques to find similarities between CV and vacancy texts. Earlier research on JRSs \cite{qin2018enhancing,qin2020enhanced,zhu2018person} predominantly makes use of convolutional neural networks (CNNs), recurrent neural networks (RNNs), and long short-term memory models (LSTMs), following the trend of most NLP-related research of the time. Recently, however, transformer architectures \cite{vaswani2017attention} have seen widespread popularity in the field of NLP due to their high-quality text embedding generation, leading to their state-of-the-art performance on many different NLP tasks \cite{he2020deberta,takase2021lessons,zhang2021poolingformer,zoph2022designing}, including job recommendation \cite{guan2024jobformer,he2021self,kaya2023exploration,lavi2021consult}.

Most pre-trained models, such as BERT \cite{devlin2018bert} and e5 \cite{wang2022text} also have a multi-lingual variant \cite{devlin2018bert,wang2024multilingual}, which has been pre-trained instead on a dataset of texts in a variety of languages. Since CVs and vacancies are susceptible to a language mismatch, being able to determine cross-lingual similarity is crucial. Based on these considerations, we formulate the following hypothesis: \textbf{H1} - \textit{The state-of-the-art models will perform substantially better than the simple baselines in terms of ranking metrics due to their higher expressive power and capability to find cross-lingual matches.}

% TODO: include more specific NLP-based JRS papers (Professional Network Matters: Connections Empower Person-Job Fit, Learning to match jobs with resumes from sparse interaction data using multi-view co-teaching network)

% We expect the transformer-based models to greatly outperform the more simplistic models, especially because bag-of-words-based techniques such as TF-IDF and Doc2Vec struggle considerably with multi-lingual data. 

\subsection{Knowledge Graph-based Approaches for Job Recommendation}
While transformer-based approaches can perform well when matching different textual features, they often entirely ignore a second type of data available for candidates and vacancies: structured data. % Recruitment agencies often have a considerable amount of structured data available for their users, such as their skill set and education. 
While parts of such data are also present in candidates' CVs, details are often omitted for brevity limiting the potential richness of the dataset. The use of \emph{knowledge graphs} presents a solution by enabling the integration of unstructured and (semi-)structured data. Furthermore, they also allow for models to be inherently explainable \cite{tiddi2022knowledge}. 

% General models such as KGAT \cite{wang2019kgat} and HAKG \cite{sha2021hierarchical} illustrate the potential of knowledge graphs in broader applications, suggesting their suitability for job recommender systems. 
We leverage these techniques to develop a knowledge graph-based JRS that performs well and is inherently explainable to users without considerable technical knowledge. While attention values provide an acceptable amount of explainability to lay users, their inability to distinguish positive and negative contributions of specific features/nodes/edges to the final prediction can make them difficult to understand fully \cite{schellingerhout2024creating}. To address this shortcoming, recent explainable graph neural network architectures have been designed to generate multiple explanations for a single graph, which can then be optimized to, for example, separate positive and negative contributions to the model's decision \cite{teufel2023megan}. This leads us to the following hypothesis: \textbf{H2} - \textit{By creating a deep attention-based heterogeneous graph neural network, we can create a model capable of generating rich multi-stakeholder embeddings, allowing it to outperform the baseline models in terms of ranking metrics. The model's node- and edge-attention values can be used as explanations.}

Furthermore, by focusing on different paths in the graph, different explanations can be generated for the same candidate-vacancy pair. This makes the use of knowledge graphs suitable for multi-stakeholder explainability. %, as different stakeholders can be provided distinct explanations, in which the paths that are more relevant for their specific requirements have been attributed more importance. 

\subsection{Fairness in Ranking}
In addition to the requirement that job recommendations need to be explainable, recent legislation also requires systems to fairly treat data subjects with different sensitive attributes \cite{euaiact2024}. While a considerable amount of research has been done on fairness in recruitment, the focus often lies on either user-sided \emph{or} provider-sided fairness, rather than on a multi-stakeholder definition of fairness \cite{alessandro2023fairness,abdollahpuri2019multistakeholder,abdollahpouri2020multistakeholder,sonboli2022multisided}. % Considering fairness can often be a balancing act, where improving fairness for one group can lead to degradation for another \cite{liu2022accuracy}, it is important to take a holistic view of the matter. 
As a result, we present the following hypothesis: \textbf{H3} - \textit{We expect model performance to be inversely correlated with both user- and provider-side fairness metrics. Considering the sparsity of positive labels (successful matches) in the dataset, achieving fairness will be difficult without deteriorating general model performance.}

We focus specifically on regional fairness and the distinction between rural and urban job seekers and providers. In the Netherlands, as in much of the world, there is an ever-growing `divide' between the highly urbanized areas and the rest of the country. This divide often leads to (parts of) the countryside feeling `left behind' and takes shape across different countries and cultures (e.g., the US, the UK, Europe, China, and South-Africa) \cite{accordino2019introduction,Baas2023,fong2009digital,Harteveld2019,knight2010rural,lembani2020same,pateman2011rural,scott2007urban}.  %Our dataset mostly contains candidates and companies residing in the Netherlands. In the Netherlands, there is an ever-growing `divide' between the highly urbanized \emph{Randstad} area and the rest of the country \cite{nijmeijer2000randstad,koster2022meer}. The Randstad, which consists of the four main urban hubs of the country: Amsterdam, Utrecht, Rotterdam, and The Hague, as well as their proximate municipalities, has been accumulating an increasingly large share of the country's wealth, population, and job opportunities \cite{CBS2011,SanderAaldersRaspe2022}. This has led to increased political tension between inhabitants of the two regions, in which those outside the Randstad feel neglected and less important to the country's political and economic landscape \cite{Baas2023,Harteveld2019}. 

\subsection{Contributions}
Our work contributes to the field by comparing the performance of various models on a high-quality dataset and introducing a novel heterogeneous knowledge graph-based JRS capable of providing both candidate- and company-sided recommendations in an explainable manner. We assess these models not only based on their predictive accuracy but also consider performance for both user- (job seeker) and provider-side (employers) fairness metrics.
\section{Methodology}
In this section, we first describe our data pre-processing pipeline. Then, we go over our novel model's architecture and describe and substantiate the baselines we used. We also provide details on how all models were trained. Lastly, we go over our approach to evaluating all the models. 

\subsection{Data}
Our explainable, multi-stakeholder job recommender system and all the baselines were trained on two separate datasets - one proprietary commercial dataset,\footnote{The company name has been anonymized} and one publicly available dataset (Zhaopin).\footnote{\url{https://www.zhaopin.com/}}
These datasets both contain a wide range of features, with varying levels of relevance for our task. For the proprietary dataset, we used the following set of features: (i) \textit{Work experience} - previous work experience of the candidates stored as tabular data; (ii) \textit{Candidate description} - user-generated texts representing the candidate's work and education history; (iii) the driver's licenses held by each candidate; (iv) the languages spoken by each candidate; (v) the skills held by each candidate; (vi) \textit{Requests} general structured data on vacancies (e.g., working hours, required education, company name) as well as vacancy texts stored as plain text.

Similarly, for the Zhaopin dataset, we made use of the following features: (i) \textit{Work experience} - unstructured `list' of skills and previous positions; (ii) the highest level of education attained by the candidate; (iii) the desired and current location (anonymized), industry, job type, and salary range of the candidate; (iv) \textit{Job description} - description and requirements of the position.
    
% In the proprietary dataset, companies make \emph{requests} to recruit additional personnel. Henceforth, we refer to requests, job descriptions, and vacancies interchangeably for simplicity. 

Both datasets contained ground truth values indicating to what extent candidates and vacancies matched each other. The labels ranged from \textit{-1} to \textit{5} in the proprietary dataset, and from \textit{0} to \textit{3} in the Zhaopin dataset. In the proprietary dataset, both -1 and 0 indicated a rejection, with -1 representing a randomly sampled rejection and 0 a manual rejection. These randomly sampled rejections were included as negative sampling, since manual rejections often still had considerable similarity to each other (i.e., they passed the initial screening). Randomly sampled rejections, by contrast, provided the model with exposure to fully irrelevant combinations. The labels 2 to 5 range from the candidate being in a try-out phase up to a successful hire. 
For the Zhaopin dataset, a 0 indicated no interaction between the two parties. On the other end, 1 through 3 indicated an interaction, application, or hire, respectively. 

% Descriptives of both datasets can be found in \cref{tab:data_descriptives}. 
Since the baseline methods we used only allowed for textual data, the input data for those was limited to the CVs and vacancies. Due to the highly sensitive nature of the proprietary dataset, it will not be shared publicly.

% \begin{table*}[!t]

% \begin{tabular}{@{}r|cccccc@{}}
% \toprule
% Dataset     & Nodes & Edges & Candidates & Vacancies & \% rural candidates & \% rural vacancies \\ \midrule
% Proprietary & 261,086   & 2,707,558 & 3,187          & 86,074        & 61.98               & 65.82                   \\ \midrule
% Zhaopin     & 276,496   & 5,570,154 & N              & N             & N/A                 & N/A \\ \bottomrule
% \end{tabular}
% \vspace{0.05in}
% \caption{The summary statistics of the final knowledge graph, as well as descriptive data. Rural candidates and rural vacancies are protected groups (relative to urban).}\
% % \vspace{-0.25in}
% \label{tab:data_descriptives}
% \end{table*}

\subsection{Knowledge Graph Creation}
\label{sec:kg_creation}
We incorporated the structured and unstructured features into a single knowledge graph. This was done in Python 3.10 using the kglab library (version 0.6.6).\footnote{\url{https://derwen.ai/docs/kgl/}} The column and table names in the datasets were converted to edge types (e.g., in a table `candidate\_skills' containing candidate IDs and skill IDs, the triples would have the form (candidate\_id, has\_skill, skill\_id)). We then created inference rules using the Web Ontology Language \cite{W3C2009OWL2} based on the relations in the full graph relating to the hierarchy of different job types (as defined by ISCO) and education levels, as well as transitive, inverse, and sub-class relationships.\footnote{The pipeline to create the knowledge graph, as well as all other code used for our experiments, can be found in our \href{https://github.com/Roan-Schellingerhout/OKRA_ECIR}{GitHub repository} (\url{https://github.com/Roan-Schellingerhout/OKRA_ECIR})}  

The full graphs were then used to create smaller sub-graphs for each candidate-vacancy pair for which we had labeled data, as well as several randomly sampled non-matching pairs. This sub-graph creation was done using a k-random walk algorithm. We opted for a random walk approach over a comprehensive one (i.e., one found through breadth-first search for example) to reduce noise in the final embeddings, as extremely distant connections can lower embedding quality \cite{sha2021hierarchical}. For the same reason, we set the maximum path length of this random search to $7$ (i.e. $k=7$) based on the findings from Sha et al. \cite{sha2021hierarchical}.

These candidate-vacancy graphs could simultaneously function as \emph{vacancy-candidate} graphs by reversing their edges. Using both the candidate-vacancy and vacancy-candidate graphs during training is what allows our model to consider the problem from a multi-stakeholder perspective, as that enables it to `ground' its reasoning from either the candidate or vacancy perspective. The actual task of our models was then to rank these sub-graphs in order of relevance, i.e., we considered the problem to be a graph-ranking task. For both datasets, we used 80\% of the candidates (i.e., all sub-graphs of 80\% of the candidates) in our training set, 10\% in our validation set, and the final 10\% in our test set. The nodes and edges within each sub-graph were stored according to a unique ID, even when referring to the same nodes in the full knowledge graphs, to prevent the models from `remembering' items in the test set. 

\subsection{Model architecture}
\begin{figure*}[!htb]
    \centering
    \includegraphics[width=\textwidth]{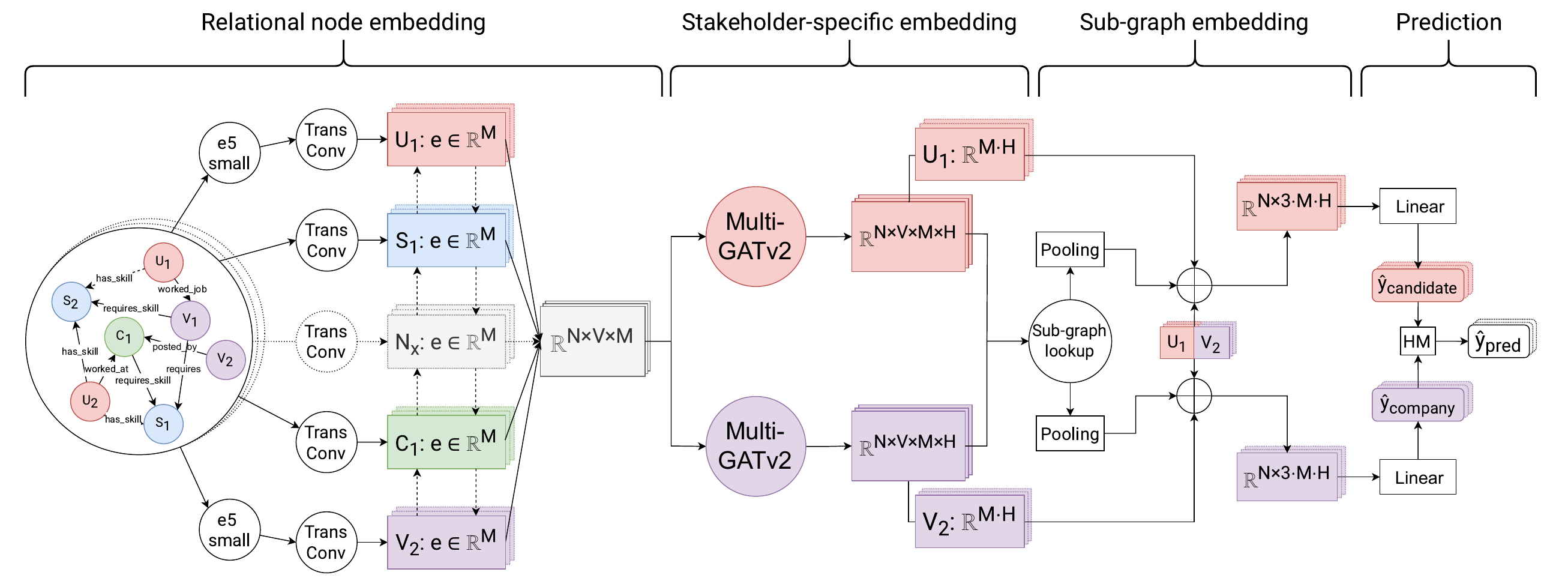}
    \caption{An overview of OKRA's architecture: the relational node embedding layer, stakeholder-specific embedding layer, sub-graph embedding layer, and prediction layer.}
    \label{fig:model}
\end{figure*}

Our novel explainable job recommender system, the \textit{Occupational Knowledge-based Recommender using Attention (OKRA)} consists of four components (\cref{fig:model}):

\begin{description}
    \item[Relational node embedding:] This part initially embeds the nodes in the sub-graph based on their own value, as well as that of the nodes in their (distant) neighborhood;
    \item[Stakeholder-specific embedding:] This component creates sets of separate candidate- and company-side node embeddings based on the candidate-vacancy and vacancy-candidate graphs;
    \item[Sub-graph embedding:] Here, the model combines the individual node embeddings into a single sub-graph embedding;
    \item[Prediction layer:] The final layer of the model calculates a candidate- and company-side matching score, whose harmonic mean is then used as the final matching score. 
\end{description}

\subsubsection{Relational node embedding}
In the first phase, the model creates node-level embeddings for each sub-graph. This is done using a combination of two approaches: nodes containing a significant amount of textual data (i.e., candidate descriptions and vacancy texts) are embedded using a transformer model.\footnote{The transformer used was \href{https://huggingface.co/intfloat/multilingual-e5-small}{e5-multilingual-small}} Due to hardware limitations, only the first 96 tokens of each text were fed into the model; we opted for simply taking the first tokens, usually containing personal descriptives and the most recent work/education history of a candidate, as that approach has been shown to perform well in previous work \cite{lavi2021consult}. The output is then mean-pooled, and fed into a linear layer to create an embedding of the desired size (this textual embedding size, $T$, is a hyperparameter). These embeddings are then considered the textual nodes' starting embeddings. Nodes without considerable textual data, on the other hand, begin with a fully random embedding. After this initial embedding, all nodes (i.e., both those related to structured and unstructured data) are embedded using a general heterogeneous graph embedding layer utilizing two sequential graph transformers \cite{shi2020masked}. 

% We opted for graph transformer convolutions % \nt{before we were talking about graph embedding layer -- how does this connect to graph transformers? where should I be looking in Figure 1} 
% in this step as they are capable of finding long-distance relations in graphs. While we narrow down the range of relations in the graph in later steps, for the initial embedding, it is important to also include long-range node relationships present in the graph,
% % \nt{what does long-range mean here given that we are restrict to 96 tokens in text. Relationships between nodes I presume but not clear here.}
% as those especially are unlikely to be stored elsewhere (e.g., in the CV/vacancy text). %\nt{but anyway 96 tokens}). 

\subsubsection{Stakeholder-specific embedding}
The output of the relational node embedding layer is a tensor containing the embedding of each node in each sub-graph, $\mathbb{R}^{N \times V \times M}$, where the embedding size $M$ is a hyperparameter. To allow the model to create multiple explanations, we use a sub-graph embedding layer based on that of Teufel et al. \cite{teufel2023megan} which uses multiple GATv2 convolutions \cite{brody2021attentive}. We use four distinct graph attention networks using four attention heads ($H = 4$) to allow both positive and negative explanations to be generated for the candidate- and company-side. We alter MEGAN's design so that it allows us to generate both a candidate- and company-side embedding, meaning that each node and edge has two unique embeddings. We multiply the candidate- and company-side embeddings by each of the four attention heads' importance weightings. Then, we concatenate the results, leaving us with two refined node-level embeddings $\mathbb{R}^{N \times V \times M \times H}$. In plain terms, we are left with $H$ node embeddings of size $M$ for all nodes ($V$) in the sub-graphs in our current batch ($N$).

\subsubsection{Sub-graph embedding}
Considering the number of nodes, $V$, per graph is not fixed, as different graphs can have a different number of nodes, we stack all nodes belonging to the same sub-graph into their respective tensor. This tensor is then pooled to create a sub-graph embedding. The pooling technique used is considered a hyperparameter (mean, max, sum). This pooling operation is done twice, once for the candidate-side embedding and once for the company-side embedding, creating two sets of sub-graph embeddings $\mathbb{R}^{N \times M \cdot H}$. Lastly, the candidate- or company-side node embeddings of the two `main nodes' of each sub-graph, i.e., the candidate and vacancy for which we need to determine the matching scores, are concatenated to these sub-graph embeddings to ensure their relevance is not diminished during the pooling operation. This leads to each sub-graph having two final embeddings, one for the candidate side and one for the company side, both having shape $\mathbb{R}^{N \times 3 \cdot M \cdot H}$. In other words, for each sub-graph in the batch, we have a single embedding consisting of the pooled, concatenated head embeddings of each node in the sub-graph, which in turn has been concatenated to the full embeddings of its two `main nodes' - the relevant candidate and vacancy nodes. 

\subsubsection{Prediction layer}
These two matrices are then fed into two separate linear layers that determine the candidate- and company-side matching score between the candidate and the vacancy in the sub-graph (a single floating point number between -100 and 100, as to be human-understandable). The harmonic mean of these two values is considered to be the final matching score of the candidate and vacancy, using which the sub-graphs are ultimately ranked. As a result, the output of the model is a single vector $\mathbb{R}^{N}$. We opted for the use of the harmonic mean over the arithmetic mean, as we considered lower values to be more impactful - if a candidate is very under-qualified for a position they could be highly interested in, it is still not useful to recommend this to them, as the odds of a successful match being made are slim. 

\paragraph{Implementation and hyperparameters} The model was implemented using a combination of the Pytorch \cite{paszke2019pytorch} and PyTorch Geometric \cite{fey2019fast} libraries. To perform optimization, we made use of the ADAM optimizer \cite{kingma2014adam}. We used a custom LambdaRANK \cite{burges2006learning} loss function to directly optimize our performance metric: the model's nDCG score. During training, hyperparameter tuning was performed through a Tree-structured Parzen Estimator for Bayesian optimization \cite{bergstra2011algorithms} using the Optuna \cite{akiba2019optuna} library. All training was performed using a single NVidia Tesla V100-SXM2-16GB GPU. After having run 32 trials, we found the optimal hyperparameters for OKRA to be: a textual embedding size $T$ of 128, a node embedding size $M$ of 32, and the use of \emph{mean}-pooling for both the candidate- and company-side. Optimal performance was achieved after training the model for 3 epochs with a learning rate of $5.283 \times 10^{-5}$.

\subsection{Baselines}
To determine the performance and fairness gain of our novel model, we compared it against multiple different baselines. These baselines differed widely in their complexity and approach, ranging from very simplistic to state-of-the-art. We used several baselines that perform well on general NLP tasks, as we expected them to perform similarly well while matching CVs and vacancies.

\textbf{Random} - To determine the general difficulty of the ranking task on our dataset, the first baseline we created was a random baseline. This baseline assigns every vacancy a random matching score for the candidate, which is then used to generate the list of recommendations. 

\textbf{TF-IDF} - The second baseline we used, was the ranking vacancies based on the cosine similarity of their term frequency-inverse document frequency (TF-IDF) \cite{sparck1972statistical} vectors to the CV. 

\textbf{Doc2Vec} - Another simple, text-based baseline we used, was Doc2Vec \cite{le2014distributed}. This method creates document embeddings by predicting words based on their neighboring words and a unique paragraph ID. The ranking is then again done using the cosine similarity between embeddings. 

\textbf{e5-multilingual} - As the first state-of-the-art baseline, we used a fine-tuned transformer model that was not specifically designed for recruitment data: e5 \cite{wang2024multilingual}.\footnote{We specifically used the \href{https://huggingface.co/intfloat/multilingual-e5-base}{`e5-multilingual-base`} model from HuggingFace} We used a Siamese architecture to feed the CV and vacancy texts into the model, and ranked vacancies using the cosine similarity of the embeddings. 

\textbf{conSultantBERT (cBERT)} - Furthermore, we used a similar fine-tuned transformer model that was specifically created to generate vacancy/CV embeddings: conSultantBERT \cite{lavi2021consult}. We then used the same method to fine-tune this model as with e5. Previous work has shown that task-adapted BERT-based models obtain SOTA performance in text-only job recommendation \cite{gong2024your}.

\textbf{GraphTransformer (GTrans)} - To ensure OKRA's difference in performance is not strictly due to it being able to access more data (i.e., both structured and unstructured), we also used a graph-based baseline. We did so with two versions, one with a single GraphTransformer layer, and one with two sequential GraphTransformers. This second model is an \textbf{ablation} of OKRA that contains only the first and last layer, causing it to not generate refined candidate- and company-side embeddings and scores. 

The baselines that required hyperparameter tuning were tuned the same way OKRA was. The SOTA models were implemented using PyTorch (Geometric).\footnote{The hyperparameters for each model can be found in our \href{https://github.com/Roan-Schellingerhout/OKRA_ECIR}{GitHub repository}.}

\subsection{Evaluation}
To determine the ranking performance of the models, we used nDCG@10, nDCG@5, and nDCG@3. Depending on the context, stakeholders can be interested in a longer or shorter list of matches; therefore, we reflect this in the evaluation. 

We also evaluated all models in terms of regional (urban vs. rural) fairness. Generally, we speak of user- and provider-side fairness \cite{burke2017multisided}; in our scenario, candidates represent the users, while the companies offering vacancies are the providers. Aligning with both parties' goals, we also use two distinct fairness metrics to evaluate the models.

For user-side fairness, we consider \emph{performance disparity} ($\Delta P$) \cite{fu2020fairness, li2021user}. This metric determines the extent to which users from a protected group are discriminated against regarding model performance. The performance disparity is calculated by subtracting the nDCG@10-score of the unprotected group from that of the protected group. A negative $\Delta P$ indicates the model performs \emph{worse} for users from the protected group, while a positive value indicates it performs \emph{better} for that group. We consider a performance disparity of 0 to be optimal. 

For provider-side fairness, conversely, we consider \emph{disparate visibility} ($\Delta V$) \cite{gomez2024moregin}. This metric determines the difference in exposure a group receives between the full dataset and the list of recommendations. In our dataset 65.82\% of vacancies stem from outside of the urbanized area; considering the urban area only makes up 15\% of the Netherlands \cite{van2009ruimtelijke}, but still represents nearly 35\% of all vacancies, the rural area can be considered the protected group. To calculate the disparate visibility, we then find the fraction of \emph{recommended} vacancies from rural areas and compare this to the fraction of such items in the full dataset. A lower value indicates the protected group shows up in the recommendations \emph{less} often than in the full dataset, indicating the model tends to discriminate against those jobs. For our use case, we determine 0 to be the optimal value. 
\section{Results}
Our results are shown in \cref{tab:metrics}. We describe these in relation to our three SQs. 

\begin{table*}[h]
\centering
\begin{tabular}[\textwidth]{@{}lrcccccc@{}}
\toprule
                                   & \textbf{}   & \textbf{nDCG@10} & \textbf{nDCG@5} & \textbf{nDCG@3} & \textbf{$\Delta P$} & \textbf{$\Delta V$} & \textbf{Time} \\ \midrule
\multirow{2}{*}{\textbf{Random}}   & Proprietary & 0.1951           & 0.1395          & 0.1080          & n/a                 & n/a                 & n/a           \\ \arrayrulecolor{lightgray} \cmidrule(l){2-8} 
                                   & Zhaopin     & 0.4490           & 0.4280          & 0.4186          & n/a                 & n/a                 & n/a           \\ \arrayrulecolor{black} \midrule
\multirow{2}{*}{\textbf{TF-IDF}}   & Proprietary & 0.3895           & 0.3002          & 0.2558          & \textbf{-0.0006}             & -0.1021             & 20s           \\ \arrayrulecolor{lightgray} \cmidrule(l){2-8} 
                                   & Zhaopin     & 0.6052           & 0.6414          & 0.6641          & n/a                 & n/a                 & 1s            \\ \arrayrulecolor{black} \midrule
\multirow{2}{*}{\textbf{D2V}}      & Proprietary & 0.3449           & 0.2331          & 0.1882          & -0.0857             & -0.0623             & 11m           \\ \arrayrulecolor{lightgray} \cmidrule(l){2-8} 
                                   & Zhaopin     & 0.4886           & 0.4919          & 0.5033          & n/a                 & n/a                 & 40s           \\ \arrayrulecolor{black} \midrule
\multirow{2}{*}{\textbf{e5 Multi}} & Proprietary\footnotemark[8] & 0.4066           & 0.2960          & 0.2325          & -0.0668             & -0.0469             & 21m           \\ \arrayrulecolor{lightgray} \cmidrule(l){2-8} 
                                   & Zhaopin     & 0.7408           & 0.7368          & 0.7279          & n/a                 & n/a                 & 70m           \\ \arrayrulecolor{black} \midrule
\multirow{2}{*}{\textbf{cBERT}}    & Proprietary\footnotemark[8] & 0.3933           & 0.2837          & 0.2289          & 0.0173              & \textbf{-0.0347}             & 25m           \\ \arrayrulecolor{lightgray} \cmidrule(l){2-8} 
                                   & Zhaopin     & 0.7265           & 0.6968          & 0.7029          & n/a                 & n/a                 & 64m              \\ \arrayrulecolor{black} \midrule
\multirow{2}{*}{\textbf{GTrans}}   & Proprietary & 0.6531           & 0.6080          & 0.5746          & -0.0459             & -0.0708             & 23m           \\ \arrayrulecolor{lightgray} \cmidrule(l){2-8} 
                                   & Zhaopin     & 0.4349           & 0.3344          & 0.2673          & n/a                 & n/a                 & 103s           \\ \arrayrulecolor{black} \midrule
\multirow{2}{*}{\makecell{\textbf{OKRA} \\ Ablation}} & Proprietary & 0.6810           & 0.6341          & 0.6122          & -0.0423             & -0.0695             & 29m           \\ \arrayrulecolor{lightgray} \cmidrule(l){2-8} 
                                   & Zhaopin     & 0.6442           & 0.5473          & 0.5096          & n/a                 & n/a                 & 138s           \\ \arrayrulecolor{black} \midrule                       
\multirow{2}{*}{\textbf{OKRA}}     & Proprietary & \textbf{0.7690}           &\textbf{ 0.7274}          & \textbf{0.7184}          & -0.0660             & -0.0598             & 35m           \\ \arrayrulecolor{lightgray} \cmidrule(l){2-8} 
                                   & Zhaopin     & \textbf{0.8531}           & \textbf{0.7970}          & \textbf{0.7883}          & n/a                 & n/a                 & 176s          \\ \arrayrulecolor{black} \bottomrule
\end{tabular}

\caption{Accuracy, fairness, and computation time for all models. For nDCG, higher is better. For the fairness metrics, a value closer to 0 is better. For training time, lower is better. Training time is per epoch for the SOTA models and total time for D2V and TF-IDF. For the Zhaopin dataset, no fairness metrics could be calculated, as the dataset did not include detailed location data. Scores in \textbf{bold} indicate best performance.}
% \vspace{-0.25in}
\label{tab:metrics}
\end{table*}

\footnotetext[8]{For the proprietary dataset, token counts were limited to 384 (down from 512), as a higher value did not improve performance, but greatly increased training time.}

%:  \textbf{SQ1.} How well do state-of-the-art baseline models perform on our unstructured dataset of CVs and vacancies?; \textbf{SQ2.} How can we develop an explainable graph neural network that is able to fully utilize heterogeneous data in order to outperform the state-of-the-art baselines?; \textbf{SQ3.} How does such a model perform regarding user- and provider-side fairness metrics compared to state-of-the-art baselines?

\subsection{Accuracy}

\paragraph{Baselines (SQ1)}
We used nDCG to measure accuracy as this measure is sensitive to rank. Considering the recommendations are intended to be used as a shortlist by humans, we only included the top 10 recommended items when determining model performance, as each recommendation will have to be manually considered. We see that the performance for even the simple models has some merit, especially for the Zhaopin dataset, and that TF-IDF performs comparably to conSultantBERT which was fine-tuned specifically on recruitment data. ConSultantBERT also is outperformed by e5 multi-lingual, in terms of both nDCG and training time. Doc2Vec performed the worst out of all models by a noticeable margin. Considering it also took considerably longer to train compared to TF-IDF, it can be regarded as the worst-performing model for this task. 
    
\paragraph{OKRA (SQ2)}
We see that graph-based models strongly outperformed the other models on the proprietary dataset, showing the added value of combined structured and unstructured data. The text-based models performed significantly better on the Zhaopin dataset than on the proprietary dataset. We believe that this is due to the more structured nature of the textual features of the Zhaopin dataset.%, indicating the semi-structured lists of skills and experience used in that dataset are more information-rich than the regular CVs available in the proprietary dataset. 
The proposed model, OKRA performs the best on both datasets, showing the benefit of generating separate embeddings for different stakeholders. However, this performance gain comes at some cost in training time compared to the other studied graph-based method.  

\subsection{User- and provider-side fairness (SQ3)}
In our data, the protected classes are the rural candidates (users) and vacancies (providers). While these both account for the majority in our dataset, they are \emph{relatively} underrepresented compared to their urban counterparts. % Considering the Randstad area makes up 15\% of the Netherlands, the fact that over a third of the candidates and vacancies are from that area shows a significant skew. 

\paragraph{User-side fairness}
Performance disparity is negative for most models, suggesting that they discriminate slightly against users in rural areas. The strongest disparity, and thus bias, is found for e5, which performs over 15\% better for candidates from urban areas than for candidates from rural areas. A notable exception can be seen for conSultantBERT, which has a slight positive disparity. 
 
\paragraph{Provider-side fairness} 
Provider-side disparate visibility is negative for all advanced models suggesting that they discriminate against providers in rural areas. This is a relative metric, i.e., the values in the table should be interpreted as the difference in the representation of rural companies in the recommendations compared to their visibility in the full dataset ($R_c = 0.6582$). 
This indicates that these models \textit{exacerbate} the biases that are already present in the data. Furthermore, we note that TF-IDF is the most unfair for providers, while conSultantBERT is the least unfair. Finally, of the two graph-based methods, OKRA is relatively less unfair than GraphTransformer.

In general, we show that automated ranking models, regardless of their underlying architecture, tend to worsen existing biases towards rural companies. The extent of this exacerbation differs per model, ranging from a decrease in visibility between 5.27\% (for conSultantBERT) and 15.51\% (for TF-IDF).
\section{Discussion and implications}
We find that our proposed model, OKRA, achieves the highest ranking performance on both datasets. Furthermore, we find that the proprietary dataset was challenging for the baseline models. On the Zhaopin dataset, all models, except the GraphTransformers, performed significantly better. 

\subsection{Baseline model accuracy}
Due to the similarity between the simple baseline models and the SOTA baseline models, we reject \textbf{H1} (\textit{state-of-the-art models will perform significantly better than simple baselines...}). This is a surprising result, as we expected TF-IDF to perform poorly due to its inability to handle multi-lingual data properly. A possible explanation for this could be that job recommendation using only textual data requires considerable nuance. Texts that would be good matches in other scenarios (e.g., two texts related to legal work) could be insufficient in recruitment (e.g., a CV describing an attorney, and a vacancy describing a legal assistant). Therefore, the SOTA models might have insufficient expressive power for this task, even when they perform well on most other text-ranking tasks. 

\subsection{OKRA's performance}
In contrast, our model significantly outperforms all the other studied models in terms of predictive performance. However, this comes at the cost of a longer training time and thus also environmental cost. We accept \textbf{H2} (\textit{an attention-based heterogeneous graph neural network will outperform the more limited baselines, while also being explainable}). While OKRA's training time is quite high, its inference time is not. To create a list of recommendations for a new candidate, their information should be added to the full knowledge graph (which can be done almost instantly), after which several sub-graphs should be created between the candidate and potential matching vacancies (which vacancies hold potential can be determined by a simpler, but faster model like TF-IDF or BM25 \cite{robertson2009probabilistic}). Generating these sub-graphs can be done in a matter of seconds. Lastly, these sub-graphs can be re-ranked by the trained model in less than a second. Therefore, the recommendation pipeline takes a few seconds to execute for a new candidate; for existing candidates, this process is nearly instant.

\paragraph{Fairness metrics}
% Our results are summarized in \cref{tab:summary}. 
Both user- and provider-fairness were comparable across all models. For the most part, models discriminated slightly towards people and companies from rural areas. As a result, we reject \textbf{H3} (\textit{fairness and performance are inversely correlated}). The one exception was conSultantBERT, which may be good for mitigating biases for under-represented groups. Although still unfairly treating rural companies, it was the \emph{least} unfair for providers as well. However, it also has poor ranking performance and limited explainability. % While it is difficult to determine exactly why conSultantBERT performs relatively well on both fairness metrics, one possibility is that its recruitment-specific model design allows it to be more nuanced when working with differences in urban/rural vocabulary. Lavi et al. \cite{lavi2021consult} show in their work that conSultantBERT performs significantly better than alternative models when it comes to handling the vocabulary mismatch between candidates and companies (even across different languages); similar behavior could cause it to better adapt to the language mismatch between rural and urban areas as well. 

% \begin{table}[]
% \centering
% \resizebox{\linewidth}{!}{%
% \begin{tabular}{@{}lllll@{}}
% \toprule
%                           & \textbf{nDCG@10} & \textbf{U-Fairness} & \textbf{P-Fairness} & \textbf{Time} \\ \midrule
% \textbf{Random}           & --                &                     &                     & --            \\
% \textbf{TF-IDF}           & --                &                     &                     & ++            \\
% \textbf{D2V}              & --                &                     &                     &               \\
% \textbf{e5}               &                   &                     &                     &               \\
% \textbf{cBERT}            &                   & ++                  & (++)                &               \\
% \textbf{gTrans}           &                   &                     &                     &               \\
% \textbf{OKRA}             & ++                &                     &                     & --            \\ \bottomrule
% \end{tabular}
% }%
% \caption{The relative evaluation of each model across the different metrics. ++ indicates a winning model,  -- indicates a losing model}
% \label{tab:summary}
% \end{table}

\subsection{Ethical concerns}
%Even when incorporating explainability and fairness into the design 
While final decisions should always be made by people to ensure accountability, the use of automated systems in the decision-making process in recruitment comes with ethical risks.  %Human recruiters cannot be expected to correct mistakes made by a recommender system. 
A list of recommended items often makes up a tiny fraction of the entire pool of possibilities. As a result, a recruiter relies on the system to have sufficient and diverse recall.
%itwith the most relevant options, as manually looking for false negatives is an impossible task. 
%Therefore, even when a recruiter can prevent some bias or injustice from arising, the system still holds considerable power over candidates and companies. 

Furthermore, by automating parts of the recruitment process, the workload of recruiters can be lowered. %Although this may initially seem like a benefit, which it definitely can be, this decrease in workload comes with the risk that companies will have a decreased incentive to hire recruiters. %This could lead to layoffs, as fewer employed recruiters could still perform the same amount of work. 
While these short-term cost-cutting measures may seem tempting for recruitment agencies, excluding human recruiters will inevitably cause issues for model performance. %As the job market evolves, so should the decision-making process within recruitment; humans are capable of adapting to the dynamic nature of the labor market, but 
Prediction models, which are trained entirely on historical data, will continuously reinforce previous behavioral patterns - regardless of potential inaccuracies, biases, or obsolescence. %As a result, recruitment agencies should be extremely wary of replacing human recruiters with automated alternatives. 

\subsection{Limitations and future work}
% One potentially limiting factor of our work, was the lack of available computing resources. The poor performance of the SOTA baselines could potentially be caused by excessive truncation of the inputs. However, this truncation was necessary for us to train the models on our hardware. Future work could evaluate these SOTA models on a similar task with more resources to ensure crucial details are not trimmed from the texts. Alternatively, future work could look into automatically shortening CVs and vacancies so that their most essential sections are retained, while removing information less relevant to the ranking process to decrease tokens per text. 

% Similarly, the current implementation of the graph-based models is not fully vectorizable, as the matrix representations of all subgraphs are combined into a single tensor, which has to then be iteratively reversed into sub-graph-specific representations. By parallelizing this step, better use could be made of available GPU resources, which could improve the graph-based models' training time.

In our approach, both stakeholders are treated with the same importance. While we find this suitable for recruitment, as neither stakeholder can be seen as more important for the industry, in some other domains it could be useful to tweak how much each stakeholder's requirements weigh in the decision. For example, when recommending items with high demand but low supply, it could be useful to adhere more to providers' wishes than customers', as losing providers due to poor model performance would be more impactful than losing customers. This could additionally be done for different user- and provider-side fairness metrics. Future work could look into different ways in which these types of unfairness can be negated, and how both types can be properly balanced. 

Finally, while OKRA is explainable in principle, its explanations have not been evaluated by the stakeholders. Although we have evaluated similar multi-stakeholder graph-based explanations in a previous user study \cite{schellingerhout2024creating}, explainability is a subjective and case-specific concept. Therefore, in future work, we plan to perform a similar, large-scale user study to evaluate the explanations generated \textit{by OKRA} by individuals from the three main stakeholder groups (job seekers, employers, and recruiters). % Such a user study will enable us to determine whether the stakeholder-specific explanations enable the end users to make better, more informed decisions. 

% \appendix
% \section{Appendix}
% All code used for this paper can be found in our 

\bibliographystyle{splncs04}
\bibliography{biblio}

\begin{thebibliography}{10}
\providecommand{\url}[1]{\texttt{#1}}
\providecommand{\urlprefix}{URL }
\providecommand{\doi}[1]{https://doi.org/#1}

\bibitem{abdollahpouri2020multistakeholder}
Abdollahpouri, H., Adomavicius, G., Burke, R., Guy, I., Jannach, D., Kamishima, T., Krasnodebski, J., Pizzato, L.: Multistakeholder recommendation: Survey and research directions. User Modeling and User-Adapted Interaction  \textbf{30}(1),  127--158 (2020)

\bibitem{abdollahpuri2019multistakeholder}
Abdollahpouri, H., Burke, R.: Multi-stakeholder recommendation and its connection to multi-sided fairness. In: RMSE Workshop at ACM RecSys 2019 (2019)

\bibitem{accordino2019introduction}
Accordino, J.: Introduction to bridging the “urban--rural divide”. State and Local Government Review  \textbf{51}(4),  217--222 (2019)

\bibitem{akiba2019optuna}
Akiba, T., Sano, S., Yanase, T., Ohta, T., Koyama, M.: Optuna: A next-generation hyperparameter optimization framework. In: Proceedings of the 25th ACM SIGKDD international conference on knowledge discovery \& data mining. pp. 2623--2631 (2019)

\bibitem{alessandro2023fairness}
Alessandro, F., Baranowska, N., Dennis, M.J., Graus, D., Hacker, P., Salvidar, J., Boregius, F.Z., Biega, A.J.: Fairness and bias in algorithmic hiring: a multidisciplinary survey. arXiv preprint arXiv:2309.13933  (2023)

\bibitem{Baas2023}
Baas, T.: Verschil tussen randstad en regio is onfatsoenlijk groot geworden. BNR Nieuwsradio (November 2023), \url{https://www.bnr.nl/nieuws/politiek/10530849/verschil-tussen-randstad-en-regio-is-onfatsoenlijk-groot-geworden}

\bibitem{bergstra2011algorithms}
Bergstra, J., Bardenet, R., Bengio, Y., K{\'e}gl, B.: Algorithms for hyper-parameter optimization. Advances in neural information processing systems  \textbf{24} (2011)

\bibitem{brody2021attentive}
Brody, S., Alon, U., Yahav, E.: How attentive are graph attention networks? arXiv preprint arXiv:2105.14491  (2021)

\bibitem{burges2006learning}
Burges, C., Ragno, R., Le, Q.: Learning to rank with nonsmooth cost functions. Advances in neural information processing systems  \textbf{19} (2006)

\bibitem{burke2017multisided}
Burke, R.: Multisided fairness for recommendation. arXiv preprint arXiv:1707.00093  (2017)

\bibitem{statista_2022}
Department, S.R.: Staffing industry: Global revenue (10 2022), \url{https://www.statista.com/statistics/624116/staffing-industry-revenue-worldwide/}

\bibitem{devlin2018bert}
Devlin, J., Chang, M.W., Lee, K., Toutanova, K.: Bert: Pre-training of deep bidirectional transformers for language understanding. arXiv preprint arXiv:1810.04805  (2018)

\bibitem{van2009ruimtelijke}
van Eck, J.R.R., van Amsterdam, H., Schuit, J., Noorman, N.: Ruimtelijke ontwikkelingen in het stedelijk gebied: dynamiek stedelijke milieus 2000-2006. Planbureau voor de Leefomgeving (2009)

\bibitem{euaiact2024}
{European Parliament}: European union artificial intelligence act. European Parliamentary Research Service Briefing (2024), \url{https://www.europarl.europa.eu/RegData/etudes/BRIE/2021/698792/EPRS_BRI(2021)698792_EN.pdf}

\bibitem{fey2019fast}
Fey, M., Lenssen, J.E.: Fast graph representation learning with pytorch geometric. arXiv preprint arXiv:1903.02428  (2019)

\bibitem{fong2009digital}
Fong, M.W.: Digital divide between urban and rural regions in china. The Electronic Journal of Information Systems in Developing Countries  \textbf{36}(1),  1--12 (2009)

\bibitem{fu2020fairness}
Fu, Z., Xian, Y., Gao, R., Zhao, J., Huang, Q., Ge, Y., Xu, S., Geng, S., Shah, C., Zhang, Y., et~al.: Fairness-aware explainable recommendation over knowledge graphs. In: Proceedings of the 43rd international ACM SIGIR conference on research and development in information retrieval. pp. 69--78 (2020)

\bibitem{gomez2024moregin}
G{\'o}mez, E., Contreras, D., Boratto, L., Salam{\'o}, M.: Moregin: Multi-objective recommendation at the global and individual levels. In: European Conference on Information Retrieval. pp. 21--38. Springer (2024)

\bibitem{gong2024your}
Gong, Z., Song, Y., Zhang, T., Wen, J.R., Zhao, D., Yan, R.: Your career path matters in person-job fit. In: Proceedings of the AAAI Conference on Artificial Intelligence. vol.~38, pp. 8427--8435 (2024)

\bibitem{guan2024jobformer}
Guan, Z., Yang, J.Q., Yang, Y., Zhu, H., Li, W., Xiong, H.: Jobformer: Skill-aware job recommendation with semantic-enhanced transformer. ACM Transactions on Knowledge Discovery from Data  (2024)

\bibitem{guo2020DeText}
Guo, W., Liu, X., Wang, S., Gao, H., Sankar, A., Yang, Z., Guo, Q., Zhang, L., Long, B., Chen, B.C., Agarwal, D.: Detext: A deep text ranking framework with bert. International Conference on Information and Knowledge Management, Proceedings pp. 2509--2516 (10 2020). \doi{10.1145/3340531.3412699}

\bibitem{Harteveld2019}
Harteveld, E.: Randstad versus de rest? de ruimte voor een politieke kloof tussen "centrum" en "periferie" in nederland (March 2019), \url{https://stukroodvlees.nl/randstad-versus-de-rest-de-ruimte-voor-een-politieke-kloof-tussen-centrum-en-periferie-in-nederland/}

\bibitem{he2021self}
He, M., Shen, D., Wang, T., Zhao, H., Zhang, Z., He, R.: Self-attentional multi-field features representation and interaction learning for person--job fit. IEEE Transactions on Computational Social Systems  \textbf{10}(1),  255--268 (2021)

\bibitem{he2020deberta}
He, P., Liu, X., Gao, J., Chen, W.: Deberta: Decoding-enhanced bert with disentangled attention. arXiv preprint arXiv:2006.03654  (2020)

\bibitem{kaya2023exploration}
Kaya, M., Bogers, T.: An exploration of sentence-pair classification for algorithmic recruiting. In: Proceedings of the 17th ACM Conference on Recommender Systems. pp. 1175--1179 (2023)

\bibitem{kingma2014adam}
Kingma, D.P., Ba, J.: Adam: A method for stochastic optimization. arXiv preprint arXiv:1412.6980  (2014)

\bibitem{knight2010rural}
Knight, J., Gunatilaka, R.: The rural--urban divide in china: Income but not happiness? The Journal of Development Studies  \textbf{46}(3),  506--534 (2010)

\bibitem{kumari2023siamese}
Kumari, T., Sharma, R., Bedi, P.: Siamese bi-directional gated recurrent units network for generating reciprocal recommendations in online job recommendation. In: International Conference on Innovative Computing and Communications. pp. 257--269. Springer (2023)

\bibitem{lavi2021consult}
Lavi, D., Medentsiy, V., Graus, D.: consultantbert: Fine-tuned siamese sentence-bert for matching jobs and job seekers (2021), \url{https://tika.apache.org/}

\bibitem{le2014distributed}
Le, Q., Mikolov, T.: Distributed representations of sentences and documents. In: International conference on machine learning. pp. 1188--1196. PMLR (2014)

\bibitem{le2019towards}
Le, R., Zhang, T., Hu, W., Zhao, D., Song, Y., Yan, R.: Towards effective and interpretable person-job fitting. International Conference on Information and Knowledge Management, Proceedings pp. 1883--1892 (11 2019). \doi{10.1145/3357384.3357949}, \url{https://doi.org/10.1145/3357384.3357949}

\bibitem{leksin2016job}
Leksin, V., Ostapets, A.: Job recommendation based on factorization machine and topic modelling. Proceedings of the Recommender Systems Challenge on - RecSys Challenge '16  (2016). \doi{10.1145/2987538}, \url{http://dx.doi.org/10.1145/2987538.2987542}

\bibitem{lembani2020same}
Lembani, R., Gunter, A., Breines, M., Dalu, M.T.B.: The same course, different access: the digital divide between urban and rural distance education students in south africa. Journal of Geography in Higher Education  \textbf{44}(1),  70--84 (2020)

\bibitem{li2021user}
Li, Y., Chen, H., Fu, Z., Ge, Y., Zhang, Y.: User-oriented fairness in recommendation. In: Proceedings of the web conference 2021. pp. 624--632 (2021)

\bibitem{paszke2019pytorch}
Paszke, A., Gross, S., Massa, F., Lerer, A., Bradbury, J., Chanan, G., Killeen, T., Lin, Z., Gimelshein, N., Antiga, L., et~al.: Pytorch: An imperative style, high-performance deep learning library. Advances in neural information processing systems  \textbf{32} (2019)

\bibitem{pateman2011rural}
Pateman, T.: Rural and urban areas: comparing lives using rural/urban classifications. Regional trends  \textbf{43},  11--86 (2011)

\bibitem{qin2018enhancing}
Qin, C., Zhu, H., Xu, T., Zhu, C., Jiang, L., Chen, E., Xiong, H.: Enhancing person-job fit for talent recruitment: An ability-aware neural network approach. In: The 41st international ACM SIGIR conference on research \& development in information retrieval. pp. 25--34 (2018)

\bibitem{qin2020enhanced}
Qin, C., Zhu, H., Xu, T., Zhu, C., Ma, C., Chen, E., Xiong, H.: An enhanced neural network approach to person-job fit in talent recruitment. ACM Transactions on Information Systems (TOIS)  \textbf{38}(2),  1--33 (2020)

\bibitem{robertson2009probabilistic}
Robertson, S., Zaragoza, H., et~al.: The probabilistic relevance framework: Bm25 and beyond. Foundations and Trends{\textregistered} in Information Retrieval  \textbf{3}(4),  333--389 (2009)

\bibitem{schellingerhout2024creating}
Schellingerhout, R., Barile, F., Tintarev, N.: Creating healthy friction: Determining stakeholder requirements of job recommendation explanations. In: ACM RecSys in HR '24 (2024)

\bibitem{scott2007urban}
Scott, A., Gilbert, A., Gelan, A.: The urban-rural divide: Myth or reality? Citeseer (2007)

\bibitem{sha2021hierarchical}
Sha, X., Sun, Z., Zhang, J.: Hierarchical attentive knowledge graph embedding for personalized recommendation. Electronic Commerce Research and Applications  \textbf{48},  101071 (2021)

\bibitem{shi2020masked}
Shi, Y., Huang, Z., Feng, S., Zhong, H., Wang, W., Sun, Y.: Masked label prediction: Unified message passing model for semi-supervised classification. arXiv preprint arXiv:2009.03509  (2020)

\bibitem{sonboli2022multisided}
Sonboli, N., Burke, R., Ekstrand, M., Mehrotra, R.: The multisided complexity of fairness in recommender systems. AI magazine  \textbf{43}(2),  164--176 (2022)

\bibitem{sparck1972statistical}
Sparck~Jones, K.: A statistical interpretation of term specificity and its application in retrieval. Journal of documentation  \textbf{28}(1),  11--21 (1972)

\bibitem{takase2021lessons}
Takase, S., Kiyono, S.: Lessons on parameter sharing across layers in transformers. arXiv preprint arXiv:2104.06022  (2021)

\bibitem{teufel2023megan}
Teufel, J., Torresi, L., Reiser, P., Friederich, P.: Megan: Multi-explanation graph attention network. In: World Conference on Explainable Artificial Intelligence. pp. 338--360. Springer (2023)

\bibitem{tiddi2022knowledge}
Tiddi, I., Schlobach, S.: Knowledge graphs as tools for explainable machine learning: A survey. Artificial Intelligence  \textbf{302},  103627 (2022)

\bibitem{Upadhyay2021}
Upadhyay, C., Abu-Rasheed, H., Weber, C., Fathi, M.: Explainable job-posting recommendations using knowledge graphs and named entity recognition. Conference Proceedings - IEEE International Conference on Systems, Man and Cybernetics pp. 3291--3296 (2021). \doi{10.1109/SMC52423.2021.9658757}

\bibitem{vaswani2017attention}
Vaswani, A., Shazeer, N., Parmar, N., Uszkoreit, J., Jones, L., Gomez, A.N., Kaiser, {\L}., Polosukhin, I.: Attention is all you need. Advances in neural information processing systems  \textbf{30} (2017)

\bibitem{wang2022text}
Wang, L., Yang, N., Huang, X., Jiao, B., Yang, L., Jiang, D., Majumder, R., Wei, F.: Text embeddings by weakly-supervised contrastive pre-training. arXiv preprint arXiv:2212.03533  (2022)

\bibitem{wang2024multilingual}
Wang, L., Yang, N., Huang, X., Yang, L., Majumder, R., Wei, F.: Multilingual e5 text embeddings: A technical report. arXiv preprint arXiv:2402.05672  (2024)

\bibitem{W3C2009OWL2}
{World Wide Web Consortium}: W3c standard facilitates information management and integration. \url{https://www.w3.org/press-releases/2009/owl2/} (2009)

\bibitem{yildirim2021bideepfm}
Y{\i}ld{\i}r{\i}m, E., Azad, P., {\"O}{\u{g}}{\"u}d{\"u}c{\"u}, {\c{S}}.G.: bideepfm: A multi-objective deep factorization machine for reciprocal recommendation. Engineering Science and Technology, an International Journal  \textbf{24}(6),  1467--1477 (2021)

\bibitem{zhang2021poolingformer}
Zhang, H., Gong, Y., Shen, Y., Li, W., Lv, J., Duan, N., Chen, W.: Poolingformer: Long document modeling with pooling attention. In: International Conference on Machine Learning. pp. 12437--12446. PMLR (2021)

\bibitem{zhu2018person}
Zhu, C., Zhu, H., Xiong, H., Ma, C., Xie, F., Ding, P., Li, P.: Person-job fit: Adapting the right talent for the right job with joint representation learning. ACM Transactions on Management Information Systems (TMIS)  \textbf{9}(3),  1--17 (2018)

\bibitem{zoph2022designing}
Zoph, B., Bello, I., Kumar, S., Du, N., Huang, Y., Dean, J., Shazeer, N., Fedus, W.: Designing effective sparse expert models. arXiv preprint arXiv:2202.08906  (2022)

\end{thebibliography}
\end{document}